\documentclass{elsart}         
\usepackage{epsfig}   
\usepackage{amsmath}
\usepackage{graphics}
\usepackage{amstext}
\usepackage{amsbsy}
\usepackage{amssymb}
\usepackage{amsfonts}
\begin{document}
\pagenumbering{arabic}

\thispagestyle{empty}
\def\A{\kern+.6ex\lower.42ex\hbox{$\scriptstyle \iota$}\kern-1.20ex a}
\def\E{\kern+.5ex\lower.42ex\hbox{$\scriptstyle \iT600ota$}\kern-1.10ex e}  
\newcommand{\Aname}[2]{#1}
\def\titlefoot#1{\centerline{\it #1}~\newline}
\renewcommand{\thefootnote}{\alph{footnote}}

\begin{frontmatter}

\begin{center}	
{\Large{\bf 
       	Measurement of Through-Going Particle \\\vspace{0.3cm} 
	Momentum By Means Of Multiple Scattering \\\vspace{0.5cm} 
	With The ICARUS T600 TPC}}
\end{center}


\author[WROCLAW]{\small A.~Ankowski},
\author[AQUILA]{\small M.~Antonello},
\author[LNGS]{\small P.~Aprili},
\author[LNGS]{\small F.~Arneodo},
\author[ETH]{\small A.~Badertscher},
\author[PADOVA]{\small B.~Baiboussinov},
\author[PADOVA]{\small M.~Baldo~Ceolin},
\author[MILANO]{\small G.~Battistoni},
\author[PAVIA]{\small P.~Benetti},
\author[PAVIA]{\small A.~Borio~di~Tigliole},
\author[PAVIA]{\small R.~Brunetti}{\small$^,$}\footnote{Present address:
INFN, Torino, Italy.},
\author[GRANADA]{\small A.~Bueno\corauthref{cor}},
\corauth[cor]{Corresponding author.
Tel: +34-958-244152; fax.: +34-958-248529.\hfill\break
{\it E-mail address:}\/ a.bueno@ugr.es.}
\author[PAVIA]{\small E.~Calligarich},
\author[NAPOLI]{\small F.~Carbonara},
\author[GRANADA]{\small M.C.~Carmona},
\author[AQUILA]{\small F.~Cavanna},
\author[CERN]{\small P.~Cennini},
\author[PADOVA]{\small S.~Centro},
\author[CESNEF]{\small A.~Cesana},
\author[UCLA]{\small D.B.~Cline},
\author[KRAKOW1]{\small K.~Cie\'slik},
\author[NAPOLI]{\small A.G.~Cocco},
\author[PAVIA]{\small C.~De~Vecchi},
\author[KRAKOW1]{\small A.~D\c abrowska},
\author[NAPOLI]{\small A.~Di~Cicco},
\author[PAVIA]{\small R.~Dolfini},
\author[NAPOLI]{\small A.~Ereditato},
\author[CERN]{\small A.~Ferrari},
\author[NAPOLI]{\small G.~Fiorillo},
\author[GRANADA]{\small D.~Garc\'\i a-Gamez},
\author[ETH]{\small Y.~Ge},
\author[PADOVA]{\small D.~Gibin},
\author[PAVIA]{\small A.~Gigli~Berzolari},
\author[MADRID]{\small I.~Gil-Botella},
\author[WROCLAW]{\small K.~Graczyk},
\author[PAVIA]{\small L.~Grandi},
\author[PADOVA]{\small A.~Guglielmi},
\author[KATOWICE]{\small J.~Holeczek},
\author[WARSZAWA1]{\small D.~Kie\l czewska}
\author[KATOWICE]{\small J.~Kisiel},
\author[WARSZAWA2]{\small T.~Koz\l owski},
\author[ETH]{\small M.~Laffranchi},
\author[WARSZAWA1]{\small J.~\L agoda},
\author[UCLA]{\small B.~Lisowski},
\author[GRANADA]{\small J.~Lozano},
\author[KRAKOW1]{\small M.~Markiewicz},
\author[GRANADA]{\small A.~Mart\'\i nez~de~la~Ossa},
\author[UCLA]{\small C.~Matthey},
\author[PAVIA]{\small F.~Mauri},
\author[GRANADA]{\small A.J.~Melgarejo},
\author[PAVIA]{\small A.~Menegolli},
\author[PADOVA]{\small G.~Meng},
\author[ETH]{\small M.~Messina},
\author[WARSZAWA2]{\small P.~Mijakowski},
\author[PAVIA]{\small C.~Montanari},
\author[MILANO]{\small S.~Muraro},
\author[GRANADA]{\small S.~Navas-Concha},
\author[WROCLAW]{\small J.~Nowak},
\author[UCLA]{\small S.~Otwinowski},
\author[LNGS]{\small O.~Palamara},
\author[LNF]{\small L.~Periale},
\author[AQUILA]{\small G.~Piano~Mortari},
\author[PAVIA]{\small A.~Piazzoli},
\author[LNF]{\small P.~Picchi},
\author[PADOVA]{\small F.~Pietropaolo},
\author[KRAKOW2]{\small W.~P\'o\l ch\l opek}, 
\author[WARSZAWA1]{\small M.~Posiada\l{}a},
\author[PAVIA]{\small M.~Prata},
\author[PAVIA]{\small M.C.~Prata},
\author[WARSZAWA2]{\small P.~Przewlocki},
\author[PAVIA]{\small A.~Rappoldi},
\author[PAVIA]{\small G.L~Raselli},
\author[WARSZAWA2]{\small E.~Rondio},
\author[PAVIA]{\small M.~Rossella},
\author[NAPOLI]{\small B.~Rossi},
\author[ETH]{\small A.~Rubbia},
\author[PAVIA]{\small C.~Rubbia},
\author[MILANO]{\small P.R.~Sala},
\author[PAVIA]{\small D.~Scannicchio},
\author[AQUILA]{\small E.~Segreto},
\author[UCLA]{\small Y.~Seo},
\author[PISA]{\small F.~Sergiampietri},
\author[WROCLAW]{\small J.~Sobczyk},
\author[KRAKOW1]{\small D.~Stefan},
\author[WARSZAWA2]{\small J.~Stepaniak},
\author[WARSZAWA3]{\small R.~Sulej},
\author[WARSZAWA2]{\small M.~Szeptycka},
\author[KRAKOW1]{\small M.~Szarska},
\author[CESNEF]{\small M.~Terrani},
\author[PADOVA]{\small F.~Varanini},
\author[PADOVA]{\small S.~Ventura},
\author[PAVIA]{\small C.~Vignoli},
\author[KRAKOW1]{\small T.~W\A cha\l{}a},
\author[UCLA]{\small H.~Wang},
\author[UCLA]{\small X.~Yang},
\author[KRAKOW1]{\small A.~Zalewska}

\address[WROCLAW]{\scriptsize Institute of Theoretical Physics, Wroc\l aw University,
                  Wroc\l aw, Poland}
\address[AQUILA]{\scriptsize Gruppo collegato INFN and Dipartimento di Fisica, Universit\`{a} dell'Aquila, 
               L'Aquila, Italy}
\address[LNGS]{\scriptsize Laboratori Nazionali del Gran Sasso (LNGS) INFN,
               Assergi, Italy}
\address[ETH]{\scriptsize Institute for Particle Physics, ETH H\"onggerberg,
              Z\"urich, Switzerland}
\address[PADOVA]{\scriptsize Dipartimento di Fisica, Universit\`{a} di Padova
              and INFN, Padova, Italy}
\address[MILANO]{\scriptsize Dipartimento di Fisica, Universit\`{a} di Milano
              and INFN, Milano, Italy}
\address[PAVIA]{\scriptsize Dipartimento di Fisica Nucleare e Teorica,
                Universit\`{a} di Pavia and INFN, via Bassi 6,
                I-27100 Pavia, Italy}
\address[GRANADA]{\scriptsize Departamento de F\'\i sica Te\'orica y del Cosmos and
                Centro Andaluz de F\'\i sica de Part\'\i culas Elementales (CAFPE),
                 Universidad de Granada, Granada, Spain}
\address[NAPOLI]{\scriptsize Dipartimento di Scienze Fisiche , Universit\`{a} Federico II di Napoli
                 and INFN, Napoli, Italy}
\address[CERN]{\scriptsize CERN, Gen\`eve, Switzerland}
\address[CESNEF]{\scriptsize Dipartimento di Ingegneria Nucleare, Politecnico di Milano
                and INFN, Milano, Italy}
\address[UCLA]{\scriptsize Department of Physics and Astronomy, University of California,
               Los Angeles, USA}
\address[KRAKOW1]{\scriptsize H. Niewodnicza\'nski Institute of Nuclear Physics,
                    Krak\'ow, Poland}
\address[MADRID]{\scriptsize CIEMAT, Departamento de Investigacion Basica, Madrid, Spain}
\address[KATOWICE]{\scriptsize Institute of Physics, University of Silesia,
                   Katowice, Poland}
\address[WARSZAWA1]{\scriptsize Institute of Experimental Physics, University of
                    Warszawa, Poland}
\address[WARSZAWA2]{\scriptsize A.So\l tan Institute for Nuclear Studies,
                    Warszawa, Poland}
\address[LNF]{\scriptsize Laboratori Nazionali di Frascati (LNF) INFN,
               Frascati, Italy}
\address[KRAKOW2]{\scriptsize Department of Electronics, AGH University of Science and Technology,
                  Krak\'ow, Poland}
\address[PISA]{\scriptsize INFN, Pisa, Italy}
\address[WARSZAWA3]{\scriptsize Institute of Radioelectronics, University of Warsaw, Warszawa, Poland}

\begin{abstract}
 The ICARUS collaboration has demonstrated, following the operation 
of a 600 ton (T600) detector at shallow depth, that the technique based on 
liquid Argon TPCs is now mature. The study of rare events, not 
contemplated in the Standard Model, can greatly benefit from the use 
of this kind of detectors. In particular, a deeper understanding of
atmospheric neutrino properties will be obtained thanks to the
unprecedented quality of the data ICARUS provides. However if we 
concentrate on the T600 performance, most of the
$\nu_\mu$ charged current sample will be partially contained, due 
to the reduced dimensions of the detector. In this article, we address
the problem of how well we can determine the kinematics of events
having partially contained tracks. 
The analysis of a large sample of atmospheric muons collected during the 
T600 test run demonstrate that, in case the recorded track is at least 
one meter long, the muon momentum can be reconstructed by an algorithm 
that measures the Multiple Coulomb Scattering along the particle's path. 
Moreover, we show that momentum resolution can be improved by a factor 
two using an algorithm based on the Kalman Filtering technique.

\end{abstract}


\end{frontmatter}

\newpage

\section{Introduction}
\label{sec:intro}

Liquid Argon TPCs are a most promising technique for the study of
several fundamental topics of particle physics: neutrino
properties, proton decay and dark matter. After an extensive R\&D programme, 
the ICARUS collaboration was able to operate at shallow depth a
detector of 600 tons (T600)~\cite{Amerio:2004}.
The successful completion of a series of technical tests has
shown that the liquid Argon technique is now 
mature. The sample of cosmic ray events recorded have provided us with a
statistically significant data set of unprecedented quality. Long muon
tracks, stopping muons, muon bundles, hadronic and electromagnetic
showers as well as low energy events have been studied and results
have been published 
elsewhere~\cite{Amoruso:2004dy,Amoruso:2004ti,Antonello:2004sx,Amoruso:2003sw,Arneodo:2003rr}. 

The T600 detector is now at the INFN Gran
Sasso Laboratory. It is being installed to be operated
underground. Its physics programme has been reviewed
in~\cite{T600phys}. Among the different non-accelerator physics topics
that can be covered with such a detector, the study of 
atmospheric neutrinos is particularly important. 
Thanks to its high granularity, ICARUS can separate, for all neutrino
species, charged current and neutral current events down to production
thresholds and free of detector biases. Therefore it provides a 
data set of very high quality (specially in the sub GeV region) 
and almost free of systematic uncertainties (which are nowadays the 
limiting factor in the study of atmospheric neutrino oscillations). 

Atmospheric $\nu_\mu$ charged current interactions will
produce prompt muons with mean energy of several hundreds of MeV. 
They will have travel paths of several meters, since minimum
ionizing particles deposit about 210 MeV per meter in Argon. 
However due to the reduced transverse dimensions of the T600 detector,
a large fraction of the recorded $\nu_\mu$ charged current 
will be partially contained. Calorimetric measurements will be 
highly inadequate to study this sort of events. 
To extract valuable kinematic information
from them, we can take advantage of the multiple Coulomb 
scattering underwent by muons as they propagate through liquid Argon.
Momentum measurement of partially contained tracks will be performed
in a way much similar to the one used in emulsion experiments. 
Together with this method, we try to apply a new technique, 
known as Kalman Filter~\cite{kalman}, in oder to obtain a more precise
measurement of the particle momenta. 

This article is organized as follows: Section~\ref{sec:T600} gives a
brief outline of the detector. In Section~\ref{sec:recons} we discuss
the technical aspects of the Monte-Carlo simulation and event
reconstruction. Section~\ref{sec:ms} describes what we called the {\it
classical} approach to momentum measurement by means of multiple
Coulomb scattering. In Sections~\ref{sec:kf} and~\ref{sec:kfmc} we briefly introduce the
Kalman Filter technique and how it can be used to obtain an estimation
of momenta for through-going particles. Section~\ref{sec:comparison} compares
the performance of the two algorithms. To evaluate the goodness of
the Kalman Filter method, Section~\ref{sec:kfrealdata}
describes the results obtained when the Kalman Filter algorithm is
applied to real data: in our case, a sample of atmospheric stopping muons. 
Conclusions are finally drawn in Section~\ref{sec:conclusions}.

\section{The 600 Ton Liquid Argon TPC}
\label{sec:T600}
The ICARUS T600 liquid Argon (LAr) detector~\cite{Amerio:2004}
consists of a large cryostat split in two identical, adjacent half-modules, 
with internal dimensions $3.6\times 3.9 \times 19.6$ m$^3$ each of them. 
Each half-module is an independent unit
housing an internal detector composed by two Time Projection Chambers (TPC),
a field shaping system, monitors, probes, and two arrays of photo-multipliers.
Externally the cryostat is surrounded by a set of thermal insulation layers.
The TPC wire read-out electronics is located on the top side of the cryostat.
The detector layout is completed by a cryogenic plant made of
a liquid Nitrogen cooling circuit to maintain uniform the LAr temperature,
and of a system of LAr purifiers.

A liquid Argon TPC detects the ionization charge
released at the passage of charged particles in the volume of LAr,
thus providing three dimensional image reconstruction and calorimetric  measurement of
ionizing events. The detector, equipped with an electronic read-out system, 
works as an ``electronic bubble chamber'' employing LAr as ionization
medium. 

 An uniform electric field applied to the medium makes the ionization electrons
drift onto the anode; thanks to the
low transverse diffusion of the ionization charge, the electron images of
ionizing tracks are preserved. Successive anode wire planes, biased at a
different potential and oriented at different angles, make possible the 
three dimensional reconstruction of the track image.

In each T600 half-module, the two identical TPCs are separated by 
a common cathode.
Each TPC consists of three parallel wire planes:
the first, facing the drift region, with horizontal wires (Induction plane);
the other two with the wires at $\pm 60^0$ from the horizontal direction
(Induction and Collection planes, respectively). The wire pitch is 3~mm.
The maximum drift path (distance between the cathode and the wire planes)
is 1.5~m and the nominal drift field 500~V/cm. The measured drift velocity 
is $1.55\pm 0.02$ mm/$\mu$s~\cite{Amoruso:2004ti}.

Each wire of the chamber is independently digitized
every 400~ns. The electronics was designed to allow continuous read-out,
digitization and independent waveform recording of signals from each wire of the TPC.
Measurement of the time when the ionizing event occurred (so called ``$t_0$ time'' of the event), 
together with the electron drift velocity information, provides the absolute position of the 
tracks along the drift coordinate. 
The $t_0$ can be determined by detection of the prompt scintillation light
produced by ionizing particles in LAr~\cite{T0}.

\section{Monte-Carlo simulation and event reconstruction}
\label{sec:recons} 

Our goal is to understand how accurately we can measure the momentum 
of charged particles that 
traverse the T600 detector without stopping. To this purpose, 
we have carried out a full simulation 
of muons that span a momentum range going from 250 MeV up to 6
GeV. This is the relevant energy range for the study of atmospheric
neutrinos. 

The FLUKA package~\cite{battistoni} has been 
used to simulate all relevant physics processes (multiple Coulomb scattering,
delta ray production, muon decay, energy loss by ionization, ...). The detector geometry
has been simulated according to the detector description given in
Section~\ref{sec:T600}. Special care was put in reproducing the same
experimental conditions met when the T600 test at shallow depth was
carried out. Therefore we included a detailed simulation of the 
noise and of the electronic response of the detector. 

The event reconstruction proceeded as follows: 
The wire output signals were used to identify hits (segment of track
whose energy is recorded by a wire). Hits are fit in order to extract
physical information (deposited charged, time and position). In a
latter step, hits are grouped into clusters. From them, we reconstruct
two dimensional tracks for each of the three wire planes (one
collection and two induction views). Finally the spatial coordinates
of the charged track (3D reconstruction) are obtained matching 
the hits of the 2D tracks previously
reconstructed. The key point for 3D reconstruction is the fact that
the drift coordinate is common to all three planes. This
redundancy allows to associate hits from different planes to a common
energy deposition. Finally 3D tracks are used to extract the information
provided by multiple Coulomb scattering, and thus get a measurement of
the track momentum. The same procedure was used to reconstruct the real data sample of
stopping muons used in Section~\ref{sec:kfrealdata}.

\begin{figure}[t]
   \begin{center}
     \begin{tabular}{cc}
       \epsfysize=6.5cm\epsfxsize=6.5cm\epsffile{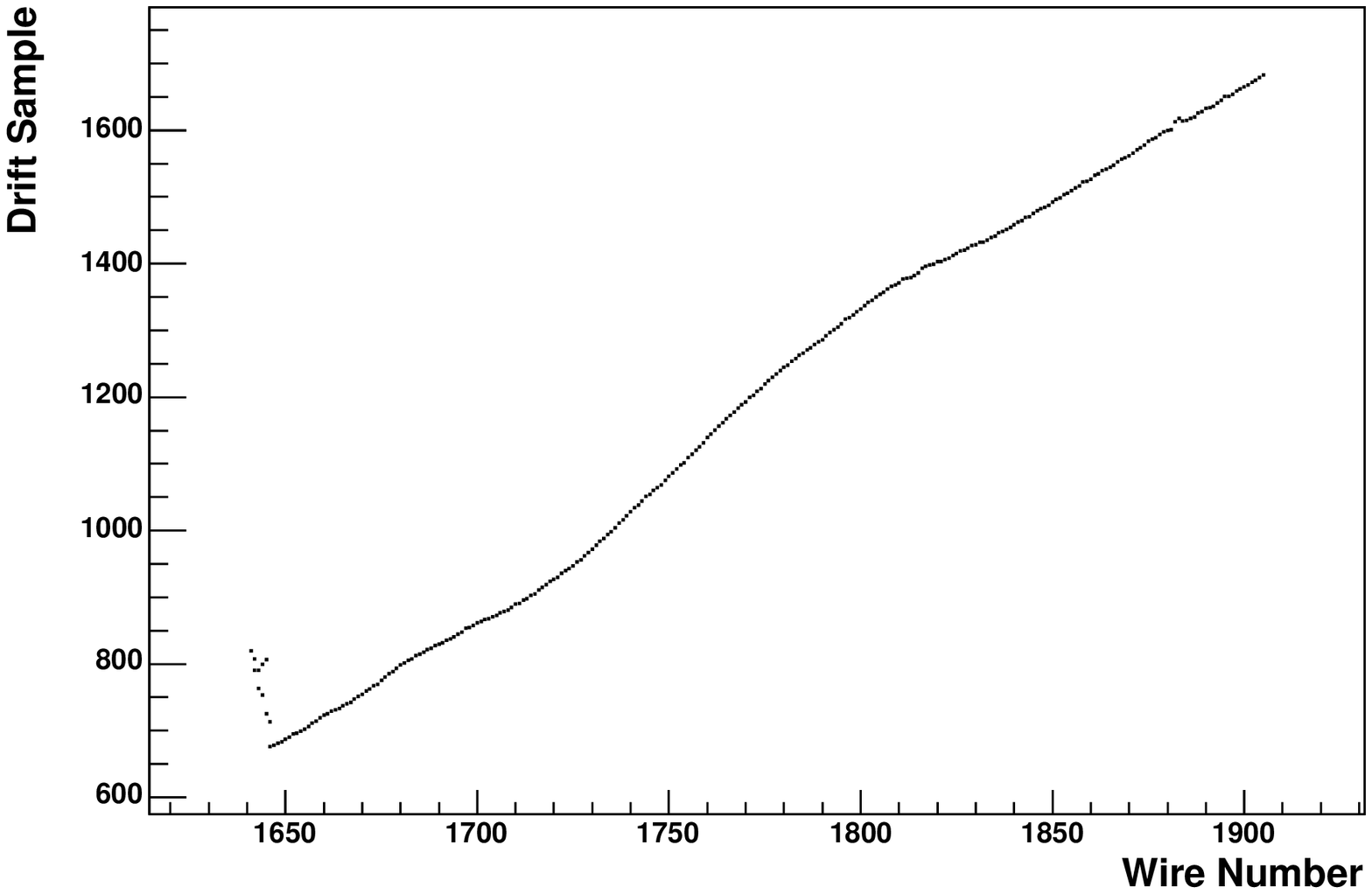} &
       \epsfysize=7.5cm\epsfxsize=7.5cm\epsffile{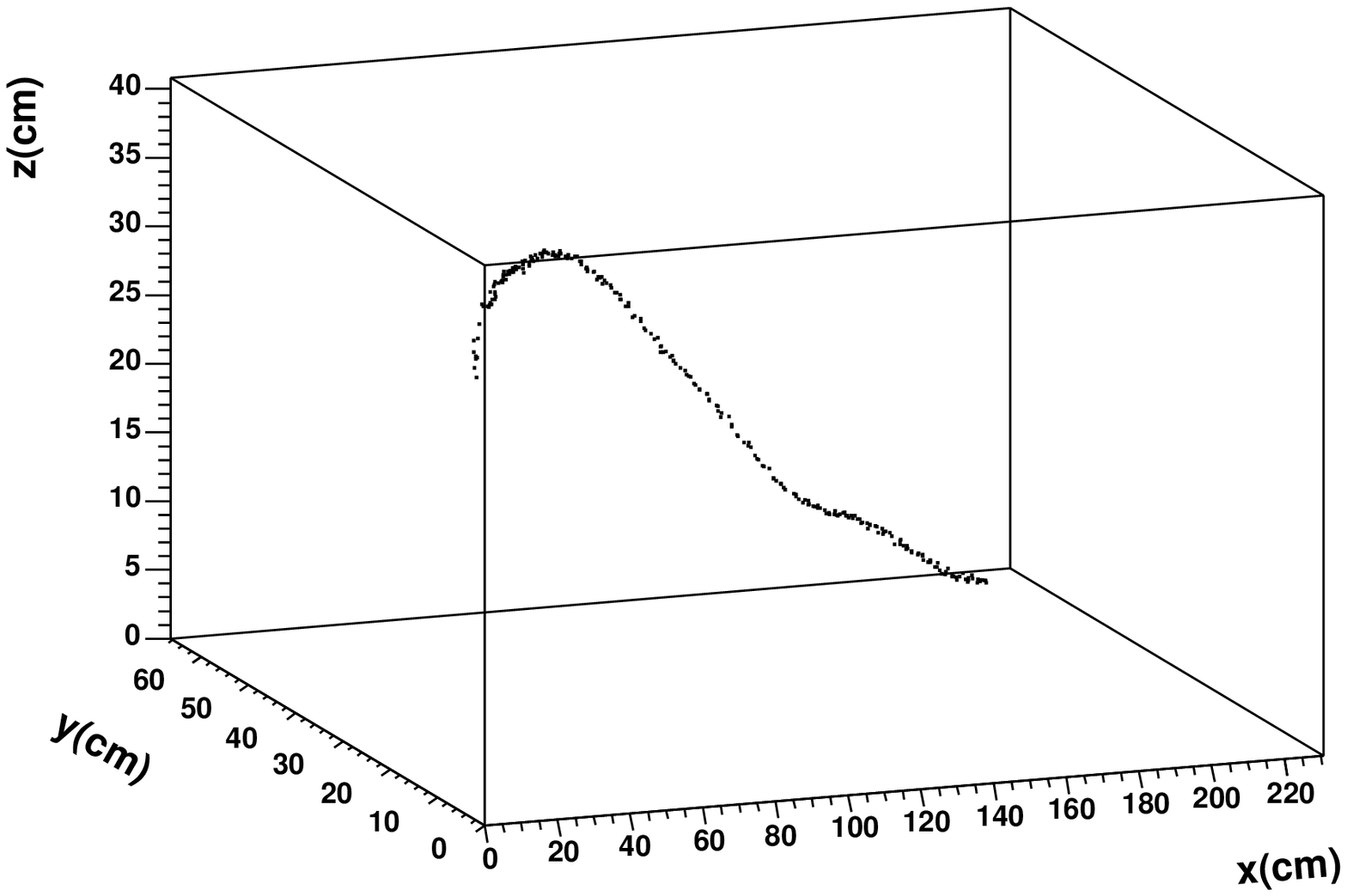} \\
     \end{tabular}
    \caption{\small
     Collection (2D) view of a typical stopping muon. 3D
reconstruction of the same muon event (right).}
    \label{fig:mcevt}
\end{center}
  \end{figure}

To exemplify how the reconstruction software works, 
figure~\ref{fig:mcevt} shows a view of a 0.5 GeV stopping muon(left panel)
and its three-dimensional reconstruction (right panel). 

\section{The {\it classical} approach to momentum measurement}
\label{sec:ms} 

We discuss what hereafter will be referred to as the 
{\it classical} approach to momentum measurement. It profits from the
fact that a charged particle traversing a medium is deflected through
many small angle scatterings, $\theta_i$. To estimate 
the momentum of a charged particle we use the following
analytical expression~\cite{Eidelman:2004wy}:

\begin{equation}
 \theta_{0}^{rms} =
 \frac{13.6\ MeV}{\beta c \; p} \,z  \, \sqrt{\frac{l}{X_{0}}}
 \left[ 1 + 0.038 \cdot ln\left(\frac{l}{X_{0}}\right)\right] 
\label{eq:theta_0}
\end{equation}

where $\theta_{0}^{rms}$ is the width of the Gaussian approximation
used for the central $98\%$ of the projected angular
distribution. $p$, $\beta$ and $z$ are the momentum, velocity and
charge of the incident particle. $X_0$ is the radiation length and $l$
the considered segment length.

The procedure used to measure the momentum of the particle is as follows: 
\begin{enumerate}

\item The whole track is split into segments of a fixed length.

\item For each segment, hits belonging to $\delta$-rays are searched
for and tagged, such that they can be excluded from the analysis. Otherwise
delta ray hits can sensibly distort the determination of the segment
direction.  

\item The remaining hits in each segment are fitted to a straight line,
 providing the segment direction.

\item For each consecutive pair of segments, the scattering angle is calculated
 as the difference between their angles.
\item We compute the RMS of the scattering angle distribution, 
after cutting out 2\% of the tails (since only the 98\% central
interval of the Moli\`{e}re distribution is Gaussian). 
\item Finally to calculate the value of
 $\theta^{rms}$, we consider all the angles but 
those whose distance to the mean is larger than 2.5 times the RMS value.
\end{enumerate}

The measured RMS of the scattering angles distribution,
$\theta_{meas}^{rms}$, is related to the ``pure'' Coulomb scattering
$\theta_{0}^{rms}$ by the following expression:

\begin{equation}
\left( \theta_{meas}^{rms}  \right)^2  =
\left( \theta_{0}^{rms}     \right)^2  +
\left( \theta_{noise}^{rms} \right)^2 
\label{eq:theta_meas}
\end{equation}

where $\theta_{noise}^{rms}$ is the angular detector resolution on the difference of two
measured segment angles. In our case, it corresponds to the spatial resolution
in the drift coordinate, $\sigma$, which is related to the error on the determination
of each individual hit time. This magnitude was measured during the T600 run,
using cosmic ray muons and test pulse data, to be about
400~$\mu$m. The noise contribution does not depend on the track
momentum. It only depends on the segment length 
($\theta_{noise}^{rms} \propto l_{seg}^{-3/2}$). Substituting in
equation~\ref{eq:theta_meas}, we get:

\begin{equation}
\theta^{rms}_{meas} = 
\sqrt{(\theta^{rms}_{0})^{2} \; + \; (\theta^{rms}_{noise})^{2}} \nonumber 
\end{equation}
\vspace{-1.cm}

\begin{equation}
 = \sqrt{\left(\frac{13.6\ MeV}{\beta c \; p} \, z \, 
\sqrt{\frac{l}{X_{0}}} \cdot 
\left[ 1+0.038\cdot ln\left(\frac{l}{X_{0}}\right)\right]\right)^{2}
\; + \; \left(C \cdot l^{-3/2}\right)^{2} }
\label{eq:conv}
\end{equation}

where $C$ is just a proportionality constant for the noise. 

To extract the track momentum, we measure $\theta^{rms}_{meas}$ 
for different segment lengths ($l$). A fit to those 
values, using equation~\ref{eq:conv}, provides an estimation 
of $p$ and $C$, which are taken as free parameters. 
This procedure allows to compute the momentum for each single track, 
since no other assumptions are made. In addition, 
with this original approach, we avoid the usual problem of choosing an
optimal segment length for the determination of the momentum.

\begin{figure}[t]
 \begin{center}
 \epsfysize=10.0cm\epsfxsize=12.0cm\epsffile{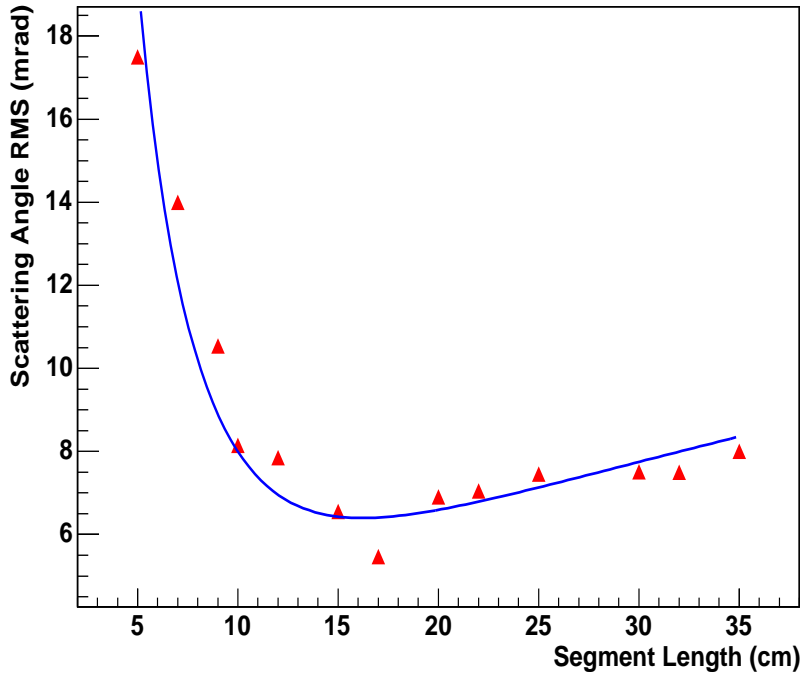} \\
 \caption{\small
  Fit of the $\theta_{rms}^{meas}$ as a function of the segment length
  for a single 3~GeV muon track.}
 \label{fig:fit}
 \end{center}
 \end{figure}

As an example, figure~\ref{fig:fit} shows the result obtained 
when this procedure is applied to a simulated 3~GeV muon. 
The triangles correspond to the experimentally measured RMS
of the scattering angles for different values of the segment length.
The curve indicates the fit result. The rising up at low values
of $l$ indicates the region where the contribution
from detector resolution dominates, whereas at high values of $l$
the main contribution comes from pure Coulomb scattering.

In this new approach, the key point to compute the momentum on 
a track by track basis is to 
decide the set of segment lengths that will be included in the
fit. The minimum segment length should be such that effects due to
multiple scattering emerge from detector noise. The optimal value for 
this minimum segment length is 5 cm. The maximum segment length 
should be short enough to allow as much entries as possible inside the angle
distribution used to compute $\theta^{rms}_{meas}$. This last value
clearly depends on the recorded track length. To improve our results, we
decided to split our sample into tracks having lengths longer than 2.5 meters
and tracks shorter than 2.5 meters. In fact, this corresponds
to a muon momentum cut at around 600 MeV. For long tracks we have 
used 13 segment lengths inside the interval [5 cm, 35 cm]. For those
tracks longer than 5 meters we have used only this length, as
using the whole track does not improve the results. For shorter 
tracks we used 10 segment lengths inside the interval [5 cm, 25 cm].

Assuming the distribution of RMS of scattering angles is Gaussian, 
it can be demonstrated that the momentum is distributed according
to the following function:

\begin{equation}
  \frac{1}{a_{0} \, p^{2}} \; 
  exp{\left({\frac{\frac{1}{p}-\frac{1}{a_{1}}}{a_{2}}}\right)^2}
\label{eq:inverse}
\end{equation}

\noindent
where $a_{1}$ gives an estimate of the momentum average.

\begin{figure}[t]
   \begin{center}
     \begin {tabular}{cc}
       \epsfysize=7.0cm\epsfxsize=7.0cm\epsffile{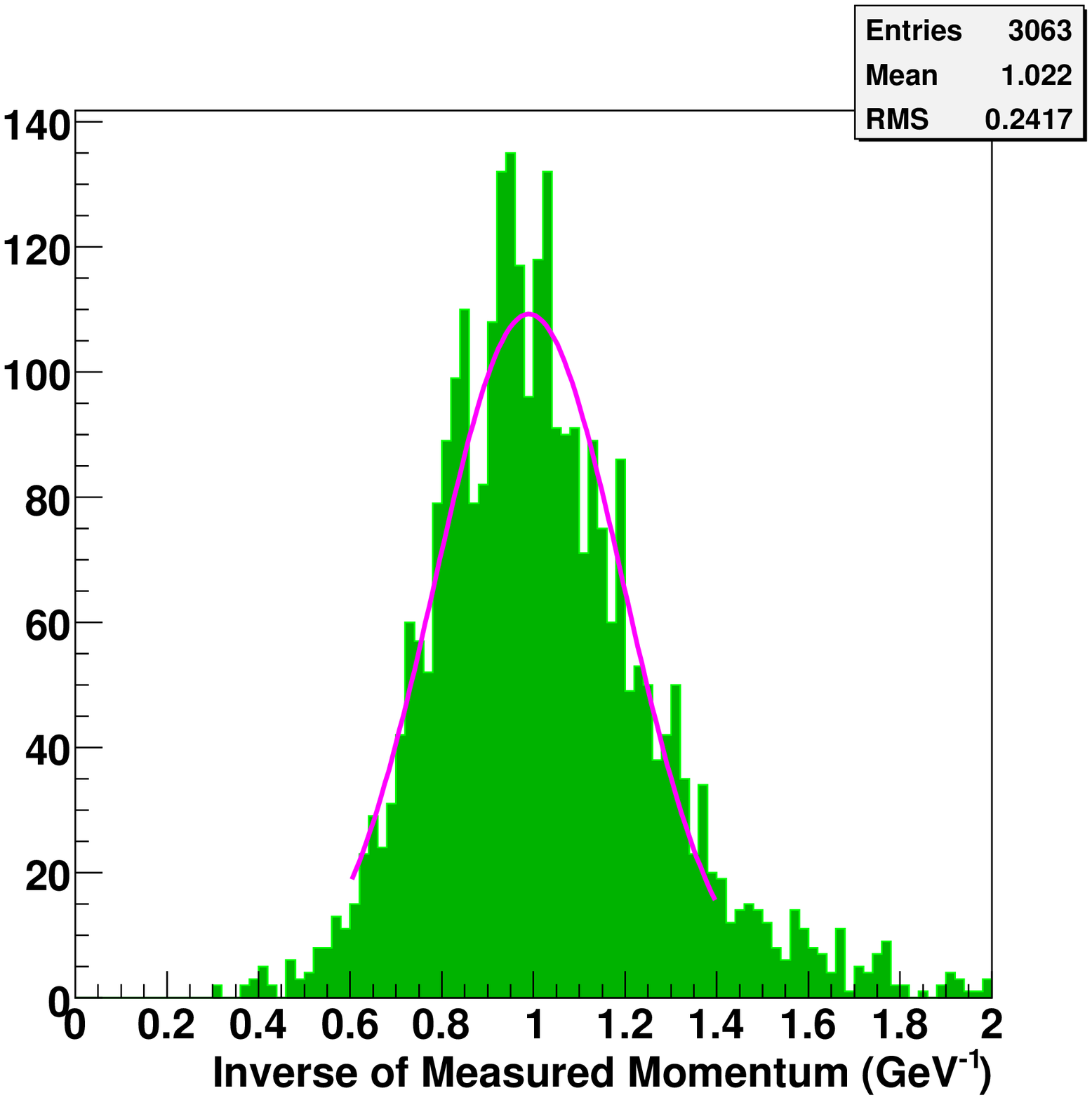} &
       \epsfysize=7.0cm\epsfxsize=7.0cm\epsffile{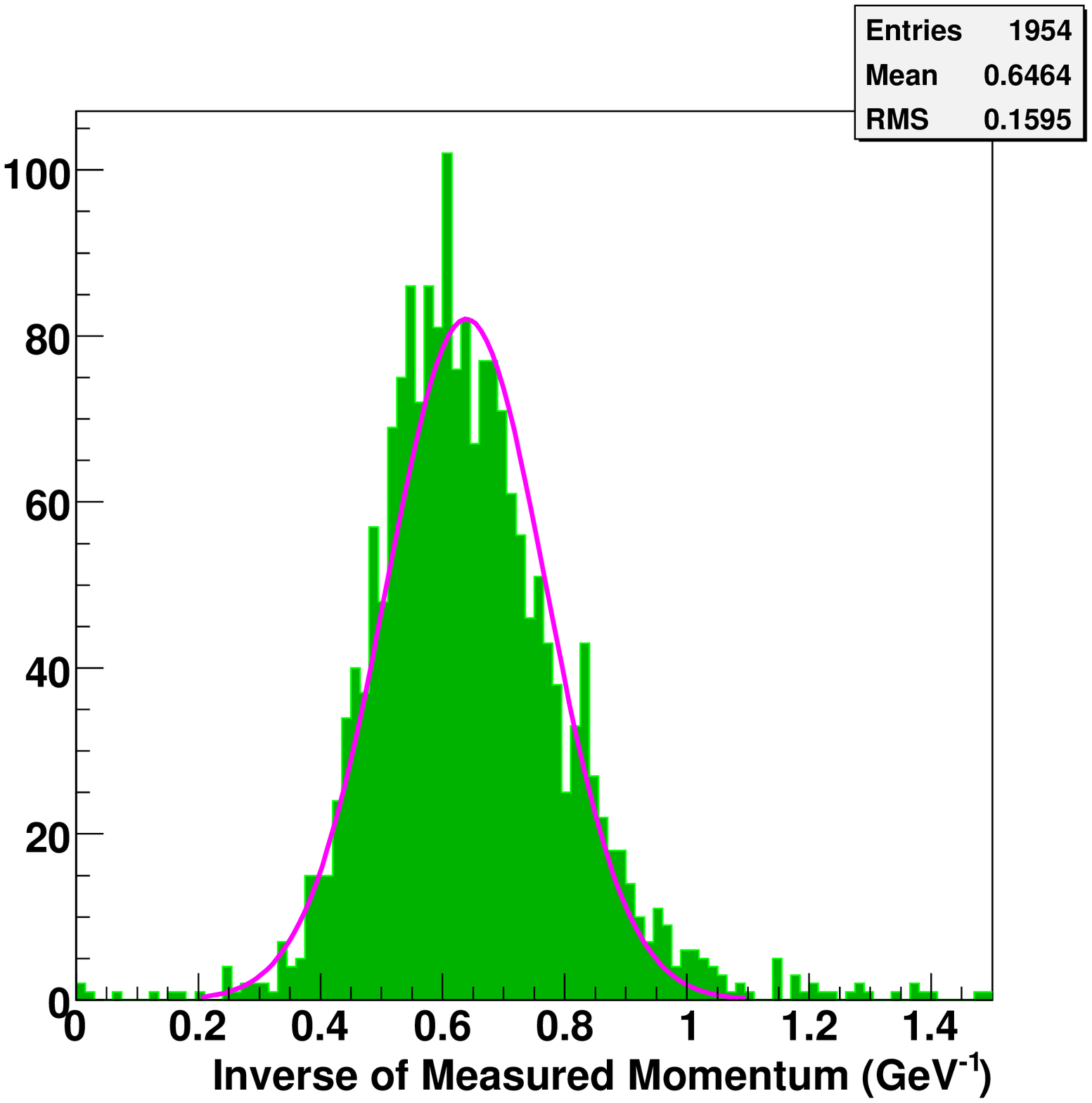} \\
     \end {tabular}
     \caption{\small Histograms of the inverse of the measured momentum for 1.5 and 2~GeV 
simulated muons.}
     \label{fig:multscatt_inverse_histo}
   \end{center}
  \end {figure}

Moreover, the momentum resolution $\Delta p$ 
is estimated from the RMS of the distribution obtained 
after computing the different scattering angle RMS, given that
{\large $\frac{\Delta\theta^{rms}}{\theta^{rms}} \equiv \frac{\Delta p}{p} \;$}. 
Figure~\ref{fig:multscatt_inverse_histo} shows the
distributions of {\large $\; \frac{1}{p} \;$} together with the Gaussian fit,
for two of the considered momenta.

The {\it classical} multiple scattering method, as described so far, 
obviously underestimates the particle momentum (see discussion 
in section~\ref{sec:comparison}). This is due to the fact 
that as the particle propagates in liquid Argon, it
loses energy by means of ionization and therefore its momentum
decreases. Since this method does not include any kind of
compensation, the final result corresponds to an average reduced momentum and
not to the momentum the particle had when it first entered the detector. 
The latter can be estimated performing an a posteriori correction since we know
the track length and the mean energy deposition of a charged particle in
liquid Argon. In the next section, we introduce a different approach, 
able to take into account energy losses along the muon track. 

\section{The Kalman Filter}
\label{sec:kf} 

The Kalman Filter~\cite{kalman} was originally proposed as 
an algorithm that deduces an optimum estimate of the past, present or future state of a dynamic 
system. It uses a time sequence of measurements of the system behaviour, information 
about its initial condition plus a statistical model that characterizes the system 
and the measurement errors. It provides a tool to separate random noises from signals
in statistical processes.

Although originally proposed as a solution for problems in communication and
signal control, the Kalman Filter has been used extensively in particle physics since it was first 
proposed as an alternative solution for track and vertex fitting and to estimate momenta~\cite{Fruhwirth:fi}. 
Typically this technique has been used in association with data provided by magnetized 
detectors. When this filter is applied, the random noise introduced in the system 
by multiple scattering can be efficiently 
separated from the track bending due to the magnetic field. Therefore an improved momentum 
measurement can be obtained.

We now explore the possibility to apply the Kalman Filter algorithm 
in an homogeneous, non-magnetized environment (the one offered by the T600) 
in order to get a measurement of the momentum of charged particles. 
The particle momentum is extracted from the information provided by the distribution 
of angles arising from multiple Coulomb scattering process. The Kalman Filter helps to disentangle 
the noise introduced by the detector resolution. In parallel, it 
takes into account the energy loss along the track, thus improving the determination of its momentum.

The Kalman Filter must be applied to a dynamic set of discrete data, therefore 
the state vector of the system should be considered only for a discrete set of points, $x_k$. 
In our case, where an homogeneous fully-sensitive detector is considered, 
those points can be chosen anywhere along the track propagation direction.
In practice, we split the track into a set of segments. The beginning and 
end points of those segments define a set of {\it planes} where we evaluate the state vector $x_k$.

The Kalman Filter is the optimal recursive estimator only in the case of a
discrete {\it linear} dynamic system. This means that all equations describing the
evolution of the state vector (covariance matrix, etc.) must be linear.
In our particular problem, the dynamic system under study (a track) is described
by the following equation:
\begin{equation}
x^-_{k} = F_{k-1} \; x_{k-1} + w_{k-1}
\end{equation}
where $F_{k-1}$ represents the {\it propagation of the state vector} from the plane k-1 to the
plane k, and $w_{k-1}$ represents the {\it noise} added to the propagation in liquid Argon,
which is of random nature and therefore smears the state vector. 
As explained below, $x_{k-1}$ and $x^-_k$ refer to the {\it filtered} and {\it predicted} state
vectors in planes {\it k-1} and {\it k}, respectively.
Two are the sources of {\it noise} we must deal with: multiple scattering and energy loss.

Usually the state vector is not directly observed.
For instance in our case the state vector contains the momentum
of the particle, which is not measured. Instead, we measure the particle scattering angle. 
At this point, a new vector called {\it measurement vector} ($m_{k}$)
must be introduced to connect the experimentally measured magnitudes with the state vector:
\begin{equation}
m_{k} = H_{k} \; x_{k} + \epsilon_{k}
\label{eq:measuredvector}
\end{equation}
where $H_{k}$ is a matrix and $\epsilon_{k}$ represents the measurement noise or
measurement errors. In general, $m_k$ and $x_k$ may have different dimensions. 

The process (system) noise $w_{k}$ is assumed to be unbiased and to have finite variance, 
being its covariance matrix $Q_k$. The measurement (detector) noise $\epsilon_{k}$ is 
assumed to be unbiased as well and to have finite variance.

\begin{equation}
\begin{tabular}{l}
$ E\left\{ w_{k} \right\} = 0 \hspace{1cm} \mathbf{Q}_{k} 
\equiv cov\left\{ w_{k}\right\}$ \\
$E\left\{ \epsilon_{k} \right\} = 0 \hspace{1.1cm} \mathbf{V}_{k} 
\equiv cov\left\{ \epsilon_{k} \right\}$ 
\end{tabular}
\end{equation}

If $w_{k}$ and $\epsilon_{k}$ are Gaussian random variables,
the Kalman Filter will be the optimal filter to the problem.

\begin{figure}[t]
   \begin{center}
    \epsfysize=11.0cm\epsfxsize=12.0cm\epsffile{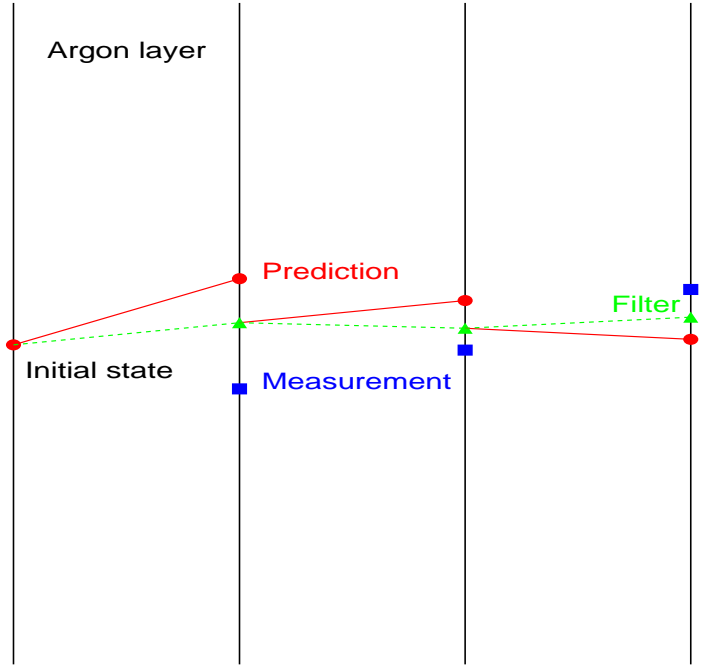} \\
    \caption{\small
      Scheme of the Kalman Filter method.
      For a given initial state, the position is propagated linearly (``Prediction'').
      Then, the ``Filtered'' state is computed by comparing the ``Measured'' and the
      ``Predicted'' states. This procedure is repeated over all track layers.
      The ``smoothing'' is done by repeating the same process backwards.}
    \label{fig:kfexample}
   \end{center}
  \end{figure}

As shown in figure~\ref{fig:kfexample}, the Kalman Filter proceeds through three conceptually
different operations\footnote{The complete set of formulae for the predicted, 
filtered and smoothed states can be found in ~\cite{Fruhwirth:fi}.}:

\begin{itemize}
\item {\bf Prediction.} Once the state vector in plane {\sl k-1} is known, the prediction consists
on estimating the state vector in a ``future'' plane {\sl k}, assuming no noise. Then,
$x_{k}^{k-1}$ (we are using the convention $x_{k}^{k-1}=x^-_k$) will be the prediction of the 
state vector at plane {\sl k}
using the corresponding state vectors up to plane {\sl k-1}.
\begin{equation}
x^-_k = F_{k-1} \; x_{k-1} 
\end{equation}

\item {\bf Filtering.} Known the predicted state vector in the plane {\sl k}, the filtering
consist in the determination of the ``present'' state vector taking into account
the information provided by all past measurements 
(through the predicted state at plane {\sl k}) and the
present measurement. If there are no measurements at a given plane, filtering does
not occur.
\begin{equation}
x_k= x^-_k + K_k (m_k - H_k \; x^-_k) 
\end{equation}
where $K$ is known as the {\it Kalman Gain} matrix and is related to the covariance of the state 
vector and the measurement noise.\\

\item {\bf Smoothing.} Once the filtering is applied to the plane {\sl k+1}, the smooth process
can be applied to estimate the state vector of a ``past'' plane {\sl k} taking into account
information up to plane {\sl k+1}.
\begin{equation}
x_{k}^n = x_k+ A_k(x_{k+1}^n-x^-_{k+1})
\end{equation}
where A is called the {\it Smoothed Gain} matrix. Here, $x_{k}^n$ refers to the final
filtered state vector after smoothing in plane {\it k}.

\end{itemize}

The power of the method lies on the fact that all previous measurements are taken into account 
to predict the future dynamical behaviour of the system. Therefore, unlike 
the case of what we called the {\it classical} approach,
where the set of reconstructed track segments is just a collection of unrelated measurements,
the Kalman Filter does take correlations into account (in particular, energy losses are
automatically included) and therefore we expect, a priori, an improved estimate 
of both the momentum and its resolution. 

\subsection{The Kalman Filter parameters}

The application of the Kalman Filter technique requires the definition of a certain set 
of vectors and matrices. The values used throughout our calculation are as follows: 

\begin{itemize}

\item[1.-] {\bf State Vector: }
To make the analysis, 
the track is split in segments of a certain length, and in each intersection
we consider several magnitudes that will conform our state vector.
In each of those segments the hits
are fitted to a straight line. The parameters of the fit will define the state vector.

In the following, we will consider a three coordinate system, with our particle traveling in the z direction, 
and being x and y perpendicular to this direction. The state vector will be made of five variables:
\begin{itemize}
\item[-]The inverse of the momentum of the particle. This magnitude
is related with the angle between the current segment and the next one
through the multiple scattering formula (see equation~\ref{eq:theta_0}).

\item[-]The position of the particle in the plane (two components: x, y).

\item[-]The slope of the track in the plane (two components: $\frac{dx}{dz}$,$\frac{dy}{dz}$). 
\end{itemize}
Hence,
\begin{equation}
{\bf x}_{k}=
\left(\begin{array}{c}
{\frac{\textstyle 1}{{\textstyle p}}}\\
x\\
y\\
\frac{\textstyle dx}{\textstyle dz}\\
\frac{\textstyle dy}{\textstyle dz}\\
\end{array}\right)
\end{equation}\\

\item[2.-] {\bf Transportation Matrix: }
The propagation matrix performs a straight line extrapolation and does not change the 
slopes, that is, assumes no interaction. The inverse of the momentum is 
propagated by subtracting from the momentum the energy lost along the segment length. 
This energy loss can be directly computed from the charge deposited in the wires by the 
particle along its path.

\begin{equation}
{\bf F}_{k}=
\left(\begin{array}{cccccc}
\frac{\textstyle 1}{\textstyle 1-\frac{\textstyle E_{lost}}{\textstyle p}} & 0 & & 0 & 0&0\\
0 & 1 & & 0 & \Delta z &0\\
0 & 0 & & 1 & 0 &\Delta z \\
0 & 0 & & 0 & 1 &0\\
0 & 0 & & 0 & 0 &1\\
\end{array}\right)
\end{equation}\\

\item[3.-] {\bf Measurement Vector and Matrix: }
The measurement vector is similar to the state vector. The only difference
is in the first row, as momentum can not be measured directly. The measured
magnitude is instead the angle difference between the incoming and outgoing trajectories, 
that is related with the momentum through equation~\ref{eq:theta_0}. This difference is taken 
into account in the measurement matrix, and hence, measurement vector and measurement matrix
are given by:

\begin{equation}
{\bf m}_{k}=
\left(\begin{array}{c}
\theta_0 \\
  x      \\
  y      \\
\frac{\textstyle dx}{\textstyle dz}\\
\frac{\textstyle dy}{\textstyle dz}\\
\end{array}\right)
, \hspace*{0.5cm}
{\bf H}_{k}=
\left(\begin{array}{ccccc}
C &  &  & &\\
 & 1 &  &  &\\
 &  & 1 & & \\
 &  &  & 1 &\\
 &  &  &  &1\\
\end{array}\right)
\end{equation}

Where $C$ is the multiple scattering constant that multiplies {\Large $\frac{1}{p}$}
in equation~\ref{eq:theta_0}.

\item[4.-]{\bf Covariance Matrices: }
The last ingredient for the method is the choice of the covariance matrices.
The covariance matrix for the system noise has been taken from reference~\cite{Wolin:1992ti} 
where the authors derive, in a simple and intuitive way, the track parameter covariance
matrix due to multiple scattering. 
They obtain all the matrix elements for two experimentally relevant track parameterizations,
being the first of them the one we are using: $x$ and $y$ slopes and intercepts.

In the covariance matrix for the detector noise, we have considered no correlations
between the components of the state vector, and hence the matrix is diagonal. 
The value for each point is just the measurement error. This amounts
to 400 $\mu$m.
\end{itemize}

\section{Test of the Kalman Filter method on a Monte-Carlo Sample}
\label{sec:kfmc} 

To assess the correctness of the choices made in the previous section in 
order to build a Kalman Filter algorithm suitable to be applied
to an homogeneous non-magnetized detector, we have simulated sets of
1000 Monte-Carlo muons for each of the following momenta:
0.25, 0.5, 1.0, 1.5, 2.0 and 3.0~GeV. 

As explained in section~\ref{sec:kf}, an initial state must be supplied for the method
to work. It can be thought, a priori, that the input guess for the
initial momentum is an important choice that may influence the final value of the momentum
measurement. In practice this is avoided by giving a big enough error to the momentum in the
first plane, such that its value becomes irrelevant.
This has been tested by carrying out a Kalman Filter with a random
initial state as input. 
The obtained momentum, close to the real one, was used as input for a second Kalman Filter. 
The final momentum does not differ significantly from the value used as input.

The only important choice for the method is the segment length to be used
in the analysis. This must be as short as possible in order
to guarantee a large enough number of points, but at the same time 
segments must be long enough in order to allow
multiple scattering effects to emerge over the detector noise. Hence a compromise must be reached.

A priori, the most ambitious idea would be to consider just single
hits, as in this way all the information is taken into account and no
average is done. But when the segment length is too short, the noise
in the measurements is too high, hiding 
multiple scattering effects. Therefore the momentum can not be properly estimated.
We have seen that to optimize our results it is advisable to use 
three different segment lengths for each track. 
For all tracks, we have considered segment lengths of 10, 12.5 and 15 cm. 
The final momentum value is obtained by averaging the momentum
computed for each of the selected segment lengths.

\begin{figure}
 \begin{center}
   \begin {tabular}{cc}
     \epsfysize=7.cm\epsfxsize=7.cm\epsffile{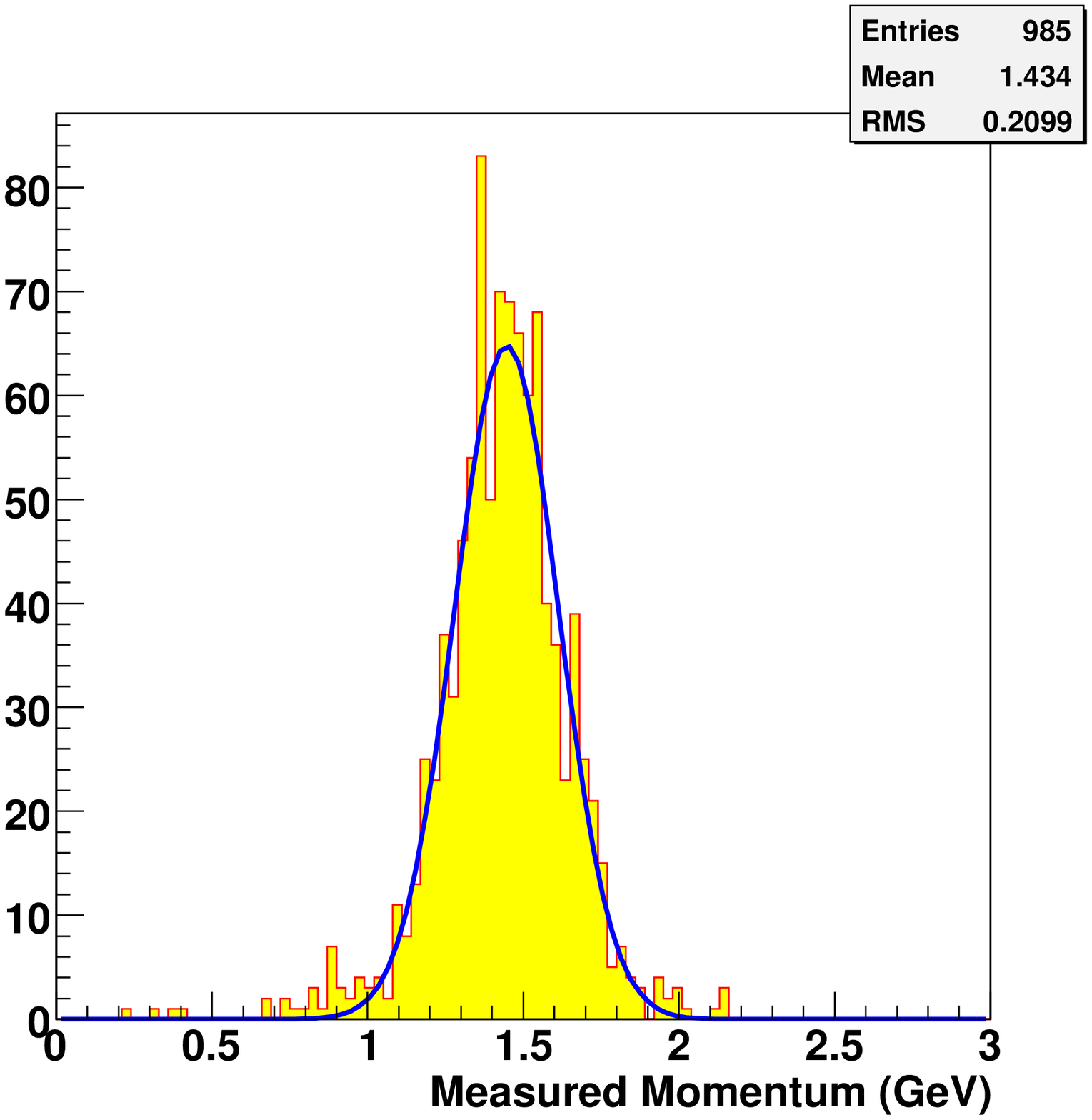} &
     \epsfysize=7.cm\epsfxsize=7.cm\epsffile{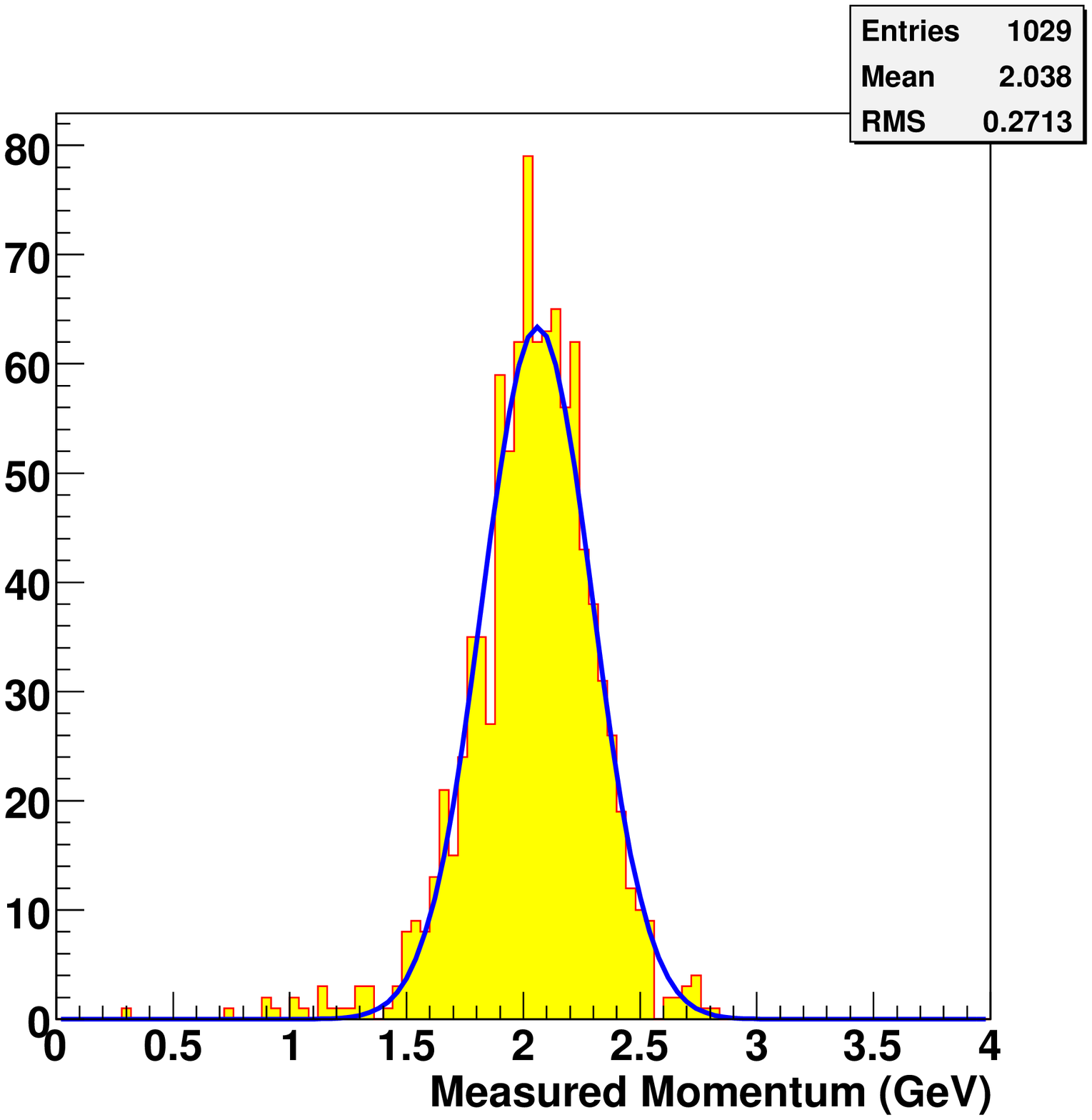} \\
   \end{tabular}
   \caption{\small Momentum distributions as given by the Kalman
Filter for simulated muons of 1.5 and 2.0 GeV.}
   \label{fig:histo}
 \end{center}
 \end{figure}

Figure~\ref{fig:histo} shows the calculated momentum distribution for two 
Monte-Carlo momenta. Each histogram entry corresponds to a reconstructed muon track. 
A Gaussian has been over-imposed to each histogram. 
The  Gaussian curve fits very well the outline of the histogram.
This clearly proves that the Kalman Filter technique allows 
to have a precise measurement of the muon momentum on a track by track basis.

We note here an intrinsic limitation of the Kalman Filter method. 
It is well known that deviations from the
original trajectory, due to multiple
scattering, become smaller as momentum increases. Detector noise starts 
playing a more important role and it 
becomes necessary to go to longer segment lengths to observe multiple scattering effects. 
Therefore for much higher momenta, a different set of segment lengths
should be used. In the limit of very high momentum, O(10 GeV), when 
multiple scattering effects are small, the Kalman Filter procedure
is not able to give a proper estimation of the particle momentum.

\section{Comparison of the performance of the {\it classical} and Kalman Filter approaches 
on a Monte-Carlo Sample}
\label{sec:comparison} 

\begin{figure}[t]
 \begin{center}
 \epsfysize=11.0cm\epsfxsize=11.0cm\epsffile{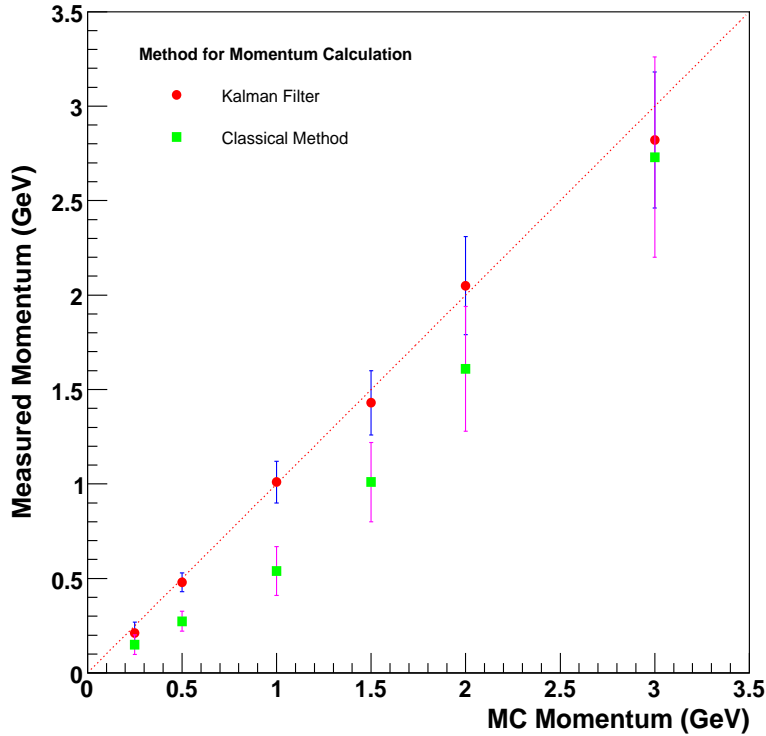} 
 \caption{\small
 Comparison between the calibration curves for {\it classical} and Kalman Filter methods.}
 \label{fig:compcal}
 \end{center}
 \end{figure}

\begin{figure}[t]
 \begin{center}
 \epsfysize=10.0cm\epsfxsize=10.0cm\epsffile{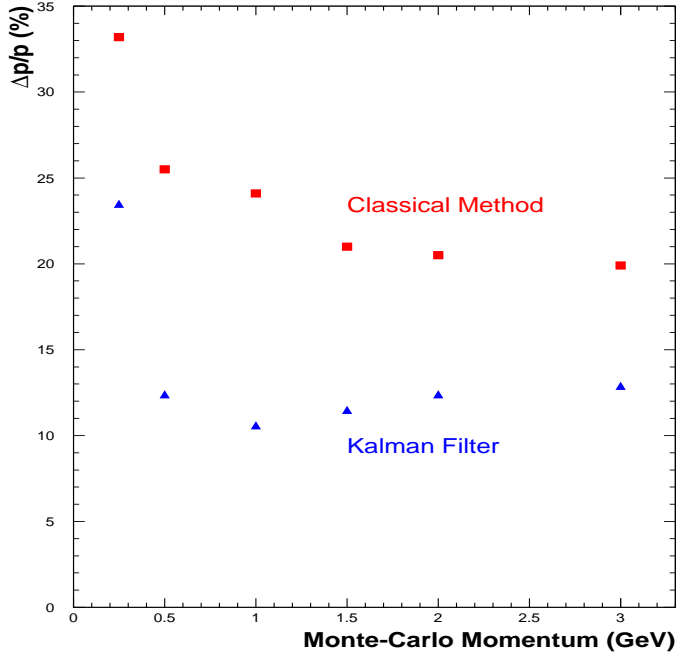} 
 \caption{\small
 Comparison between the resolutions for {\it classical} and Kalman Filter methods.}
 \label{fig:compres}
 \end{center}
 \end{figure}

We have discussed separately two algorithms aiming at measuring the
momentum of partially contained tracks. To summarize the most relevant
points, we now compare the performance of both methods. The comparison
is based on Monte-Carlo data. Figures~\ref{fig:compcal} and \ref{fig:compres} 
show the calibration curve and the 
resolutions obtained using the two methods.

In figure~\ref{fig:compcal}, the solid line shows the case where
$P_{measured}=P_{Monte-Carlo}$. The points from the {\it classical} analysis are
clearly below the line because of the energy lost by the 
muon along its path. For this approach, the error bars correspond to the RMS of
each single distribution (similar to those shown in 
figure~\ref{fig:multscatt_inverse_histo}). 

The Kalman Filter gives a much better
estimation, since it takes into account 
energy losses. For this algorithm, the error is taken as the RMS of the Gaussian fit
to the individual momentum distributions similar to those of figure~\ref{fig:histo}.
Within one standard deviation, the Kalman Filter does not
underestimate muon momentum.
Therefore, contrary to the {\it classical} method where we calculate 
the mean momentum along the track, with the Kalman Filter the momentum
at each plane can be 
calculated, and in particular, we can get a measurement of the 
most relevant magnitude, namely: the momentum at the initial plane.

Figure~\ref{fig:compres} shows the momentum resolution. For the {\it classical} 
approach it is on the range 25--20 \%. 
There is an increase of the resolution at the edges:
for small momenta, where the range of lengths and the number 
of segments is small, and for high momenta, where the effect 
of the multiple scattering decreases.

With the Kalman Filter technique, the resolutions are 
around 10\% for the whole range of studied momenta, except at
250~MeV where the resolution is close to 25\%. This increase
is easily understandable given the short track lengths involved. 
A 250~MeV muon has a typical length of less than 
1 meter. The amount of planes in which it can be split is small. Hence
the method does not have enough information to provide an accurate measurement. 
In spite of that, 250~MeV events are typically 
fully contained and measured by calorimetry with a 
very good resolution. The inclusion of this energy in the analysis was
motivated by the aim to find the limits of the method.

\begin{figure}[t]
 \begin{center}
\epsfysize=11.0cm\epsfxsize=11.0cm\epsffile{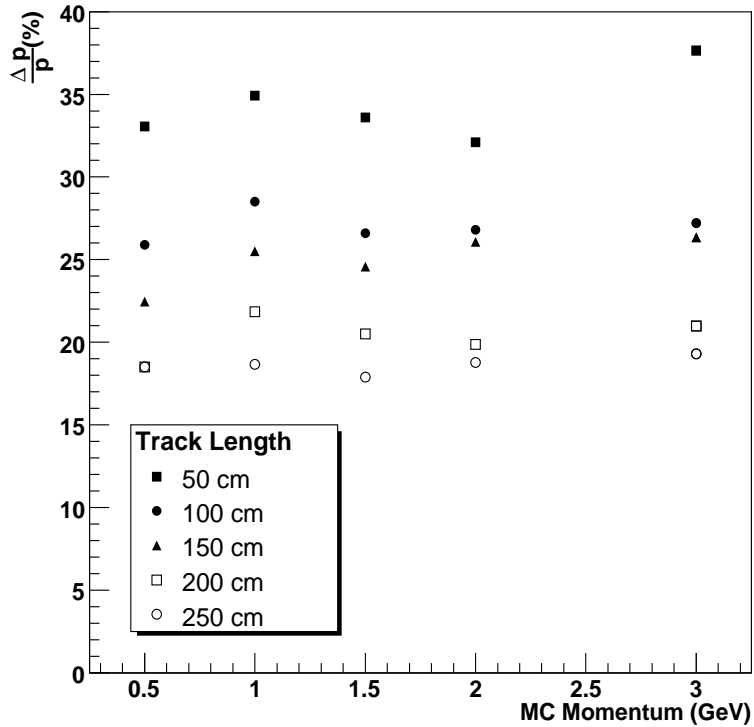} 	
 \caption{\small
 Kalman Filter algorithm: 
Monte-Carlo momentum resolution as a function of the recorded track length.}
 \label{fig:kflength}
 \end{center}
 \end{figure}

We have also studied how the momentum resolution gets affected as the
recorded track length gets shorter. This is specially important 
for the case of several GeV muons. Figure~\ref{fig:kflength} shows, for the range of
considered momenta, the expected resolutions when recorded tracks are
50, 100, 150, 200 and 250 cm long. We observed that, for all lengths, 
the momentum scale is not underestimated on average. Resolutions keep
themselves in the interval 20--25$\%$. Only for the shortest
considered tracks (50 cm), the resolution worsens up to
30$\%$. We conclude that, for tracks of a few GeV, 
the Kalman Filter allows to have a reasonable
estimate of the charged particle momenta even if
the recorded tracks are no longer than a meter.

In conclusion, with the {\it classical} approach we expect 
resolutions in the range 25--20 \% and a underestimation of
the momentum value. The results 
obtained with the Kalman Filter do not underestimate momentum and 
improve the resolution by almost a factor two. In addition, 
we notice that both methods exhibit a similar behavior
at low energies, with an increase in the error with respect to the almost constant one for
the rest of energies. This behavior appears at 500~MeV for the classical method and only at
250~MeV for the Kalman Filter. In both cases the error increase has to do with the
length of the tracks and the small amount of available segments.

\section{Kalman Filter application to a set of stopping atmospheric muons}
\label{sec:kfrealdata} 

So far, we have just relied on Monte-Carlo data to show that the Kalman
Filter is an optimal tool to estimate the momentum of partially
contained tracks. We now try to assess the goodness of this algorithm by
applying it to real data. The sample we take as reference corresponds to
a set of stopping atmospheric muons collected on the summer
of 2001 during a T600 technical run on surface. The initial set 
contains the 2690 events used in~\cite{Amoruso:2003sw}. We impose 
the condition that, in the collection view, muons should have at least 60 
hits and a minimum track length of 50 cm. These requirements leave us 
with a final sample of 1009 muons. 

Given the characteristics of the selected muons, the range of considered momenta
spans from about 200 MeV up to 800 MeV (being the mean value 400 MeV). 
For higher momenta, the sample
of recorded muons is not statistically significant. The Monte-Carlo tells 
us that the Kalman Filter performance for the lower momentum range 
is not as good as the one expected for muons above 1 GeV. 
However a demonstration that the
method performs reasonably well in the low momentum range, will
give us confidence on the fact that the algorithm can be 
also applied at higher momenta.

The selected data have been reconstructed following the procedure outlined in
Section~\ref{sec:recons}. Being fully contained, the momentum of those
muons can be very accurately measured using a calorimetric
approach. When computing momentum by means of calorimetry, recombination effects
should be taken into account. Equation~\ref{eq:calor} relates measured
and deposited charges, being R the electron recombination factor in liquid Argon:

\begin{equation}
\label{eq:calor}
Q_{meas}=R\cdot Q_{dep}\cdot e^{\frac{t-t_0}{\tau}}
\end{equation}

We take from~\cite{Amoruso:2004dy} the value $R=0.71$. 
$\tau$ stands for the drift mean life of the
electrons and its value is taken from~\cite{Amoruso:2004ti}. 
The error associated to $\tau$
is propagated to the momentum measured by calorimetry. It amounts 
to less than five per cent. 

\begin{figure}[t]
 \begin{center}
 \epsfysize=10.0cm\epsfxsize=10.0cm\epsffile{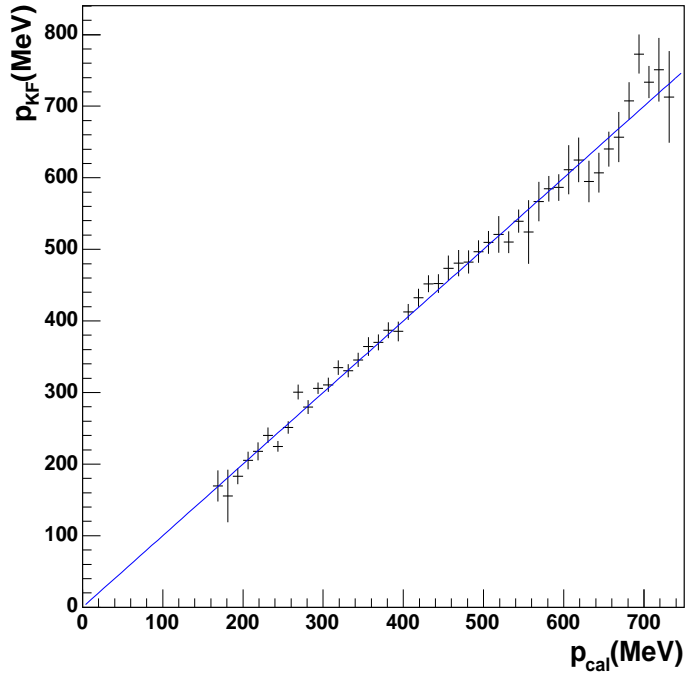} 
 \caption{\small
 Profile plot of the correlation between the momentum measured by
calorimetry $p_{cal}$ and the one obtained by means of a Kalman
Filter, $p_{kf}$.}
 \label{fig:profile}
 \end{center}
 \end{figure}

\begin{figure}[t]
 \begin{center}
 \epsfysize=10.0cm\epsfxsize=10.0cm\epsffile{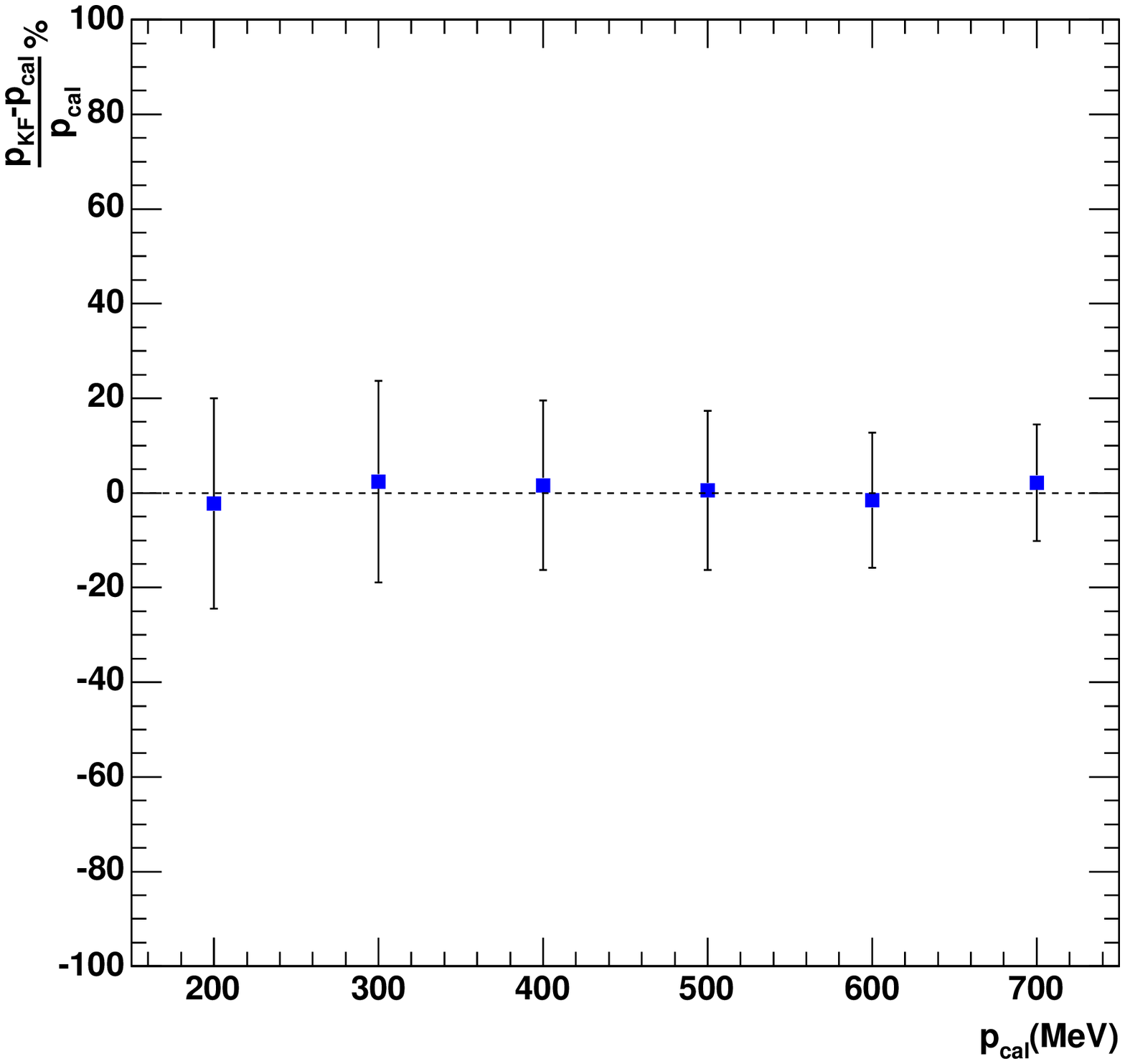} 
 \caption{\small Dispersion of the Kalman Filter measurements with
respect to the momentum measured by calorimetry, $p_{cal}$. The filled squares 
represent the mean value of the dispersion distribution. The error bars correspond 
to the RMS.}
 \label{fig:disper}
 \end{center}
 \end{figure}

In our approach we consider the momentum measured by calorimetry, 
$p_{cal}$, as the {\it true} momentum value. On a track by track
basis, we compare the momentum obtained with 
the Kalman Filter, $p_{kf}$, to $p_{cal}$. The results are shown in
figures~\ref{fig:profile} and~\ref{fig:disper}.

We plot the profile of the analyzed data set
in figure~\ref{fig:profile}. It shows that, on average, our measurements 
tend to cluster around the solid line, which depicts the ideal 
condition where $p_{kf}=p_{cal}$. In the high momentum region, 
the decrease of statistics results in bigger fluctuations 
around the central line. From the plot we infer that the relationship 
among the two performed measurements is clearly linear,
thus showing that the Kalman Filter gives a good measurement of real
data momentum.

We also plot the dispersion of the Kalman Filter measurements with 
respect to $p_{cal}$ in figure~\ref{fig:disper}. For each of the six momentum 
bins into which we group the data, the mean dispersion is very close to zero. 
The error bars correspond to the RMS of each individual distribution. For the lowest 
momenta, the error is around 20$\%$. As the momentum increases, the error bar diminishes 
reaching a minimum value of about 12$\%$. These results are in very good agreement 
with the resolutions obtained, for the low momentum range, 
in our Monte-Carlo simulation (see figure~\ref{fig:compres}).  
In view of these results, we conclude
that real data confirm the fact that the Kalman Filter algorithm, when combined
with multiple Coulomb scattering information, is an optimal
method to obtain an accurate measurement of partially contained track momenta.

\section{Conclusions}
\label{sec:conclusions} 

We have addressed the problem of measuring the momentum of partially
contained charged particles recorded with the 600 ton ICARUS detector.
This is a fundamental issue in our effort to study very precisely atmospheric
neutrinos. 

We have analyzed the performance of two independent approaches: 
\begin{itemize}
\item The one referred to as {\it
classical} extracts the momentum from the distribution of
scattering angles measured along the particle trajectory. We have
proposed an algorithm that avoids defining an optimized
segment length for the computation of the scattering angles. 
\item The second method is based on a Kalman Filter. We  have
explored, for the first time, the possibility to apply this algorithm 
in an homogeneous, non-magnetized environment. This method takes 
all measurements into account to predict the future dynamical
behaviour of the considered system. Therefore, unlike 
the {\it classical} approach,
where the set of reconstructed track segments is just a collection of unrelated measurements,
the Kalman Filter does take correlations into account (in particular, energy losses are
automatically included). 
\end{itemize}
With the {\it classical} method, the momentum is underestimated (since
energy losses are not considered). The expected resolutions lie around
25$\%$ for a range of momenta ranging from 250 MeV up to 3 GeV. With
the Kalman Filter, momentum is not underestimated and the
expected resolutions improve by a factor two those obtained with the
{\it classical} method. We have also observed that, for tracks with
momentum in the vicinity of 1 GeV or higher, the Kalman Filter provides a reasonable
momentum measurement, even in the case the recorded tracks are around
one meter long. 

To assess the goodness of the Kalman Filter approach, we have used a
set of real data composed of atmospheric muons that stop inside the
detector. The momentum of those muons is very precisely measured by calorimetry. The
momentum estimation provided by the Kalman Filter shows a very good agreement
with calorimetric measurements, proving the validity of the method. 
We therefore conclude that
the Kalman Filter approach is an excellent tool to get precise
kinematics information of the set of partially contained atmospheric
neutrino events that we expect to record at the Gran Sasso underground
laboratory.

\section{Acknowledgments}

We would like to warmly thank the many technical collaborators
that contributed to the construction of the detector and to its
operation. We are glad of the financial and technical support of
our funding agencies and in particular of the Istituto Nazionale di Fisica
Nucleare (INFN), of ETH Z\"urich and of the Fonds National Suisse de la Recherche
Scientifique, Switzerland. This work has been supported by the CICYT Grant 
FPA2002-01835. The Polish groups acknowledge the support of the State Committee 
for Scientific Research in Poland, 105,160,620,621/E-344,E-340,
E-77,E-78/SPS/ICARUS/P-03/DZ211-214/2003-2005
and 1P03B 041 30; the INFN, FAI program; the EU Commission,
TA-DUSL-P2004-08-LNGS.

\end{document}